**Synthesis of a mixed-valent tin nitride and considerations of its possible crystal structures**

Christopher M. Caskey,[1,2,3] Aaron Holder,[1] Sarah Shulda,[2] Steve Christensen,[1] David Diercks,[2] Craig P. Schwartz,[4] David Biagioni,[1] Dennis Nordlund,[4] Alon Kukliansky,[5] Amir Natan,[5] David Prendergast,[6] Bernardo Orvananos,[7] Wenhao Sun,[6] Xiuwen Zhang,[8] Gerbrand Ceder,[7] David S. Ginley,[1] William Tumas,[1] John D. Perkins,[1] Vladan Stevanovic,[1,2] Svitlana Pylypenko,[2] Stephan Lany,[1] Ryan M. Richards,[2] Andriy Zakutayev[1]*

    1) National Renewable Energy Laboratory, Golden, CO, USA
    2) Colorado School of Mines, Golden, CO, USA
    3) Larix Chemical Science, Golden, CO, USA
    4) SLAC National Accelerator Laboratory, Menlo Park, CA, USA
    5) Tel Aviv University, Israel
    6) Lawrence Berkeley National Laboratory, Berkley, CA, USA
    7) Massachusetts Institute of Technology, Cambridge, MA, USA
    8) University of Colorado, Boulder, CO, USA

*\* e-mail: andriy.zakutayev@nrel.gov*

**Abstract**

Recent advances in theoretical structure prediction methods and high-throughput computational techniques are revolutionizing experimental discovery of the thermodynamically stable inorganic materials. Metastable materials represent a new frontier for studies, since even simple binary non ground state compounds of common elements may be awaiting discovery. However, there are significant research challenges related to non-equilibrium thin film synthesis and crystal structure predictions, such as small strained crystals in the experimental samples and energy minimization based theoretical algorithms. Here we report on experimental synthesis and characterization, as well as theoretical first-principles calculations of a previously unreported mixed-valent binary tin nitride. Thin film experiments indicate that this novel material is N-deficient SnN with tin in the mixed II/IV valence state and a small low-symmetry unit cell. Theoretical calculations suggest that the most likely crystal structure has the space group 2 (SG2) related to the distorted delafossite (SG166), which is nearly 0.1 eV/atom above the ground state SnN polymorph. This observation is rationalized by the structural similarity of the SnN distorted delafossite to the chemically related $Sn_3N_4$ spinel compound, which provides a fresh scientific insight into the reasons for growth of polymorphs of the metastable material. In addition to reporting on the discovery of the simple binary SnN compound, this paper illustrates a possible way of combining a wide range of advanced characterization techniques with the first-principle property calculation methods, to elucidate the most likely crystal structure of the previously unreported metastable materials.

**I. Introduction**

The fields of solid-state chemistry and materials science are searching for and discovering new functional materials. Recently, the theoretical prediction and experimental realization of thermodynamically stable materials have seen much research[1,2,3,4] and some success.[5,6] Translating this progress to metastable materials systems, such as thermochemically unstable compounds produced by non-equilibrium thin film synthesis techniques, is an emerging frontier. For this vast metastable materials space, productive scientific approaches are lacking, because both theory and experiment face challenges here. Theoretical structure search methods are most mature when targeting ground state structures using energy as the search metric. Similarly problematic, metastable materials are likely to be initially observed in small, strained crystals that present challenges for experimental structure determination methods, such as diffraction. The opportunity, on the other hand, is huge: for every ground-state structure, there are many hundreds of higher-energy structures that may be metastable. Also, these higher-energy polymorphs may have useful practical applications, such as (anti)ferroelectric materials,[7,8] or topological insulators.[9,10]

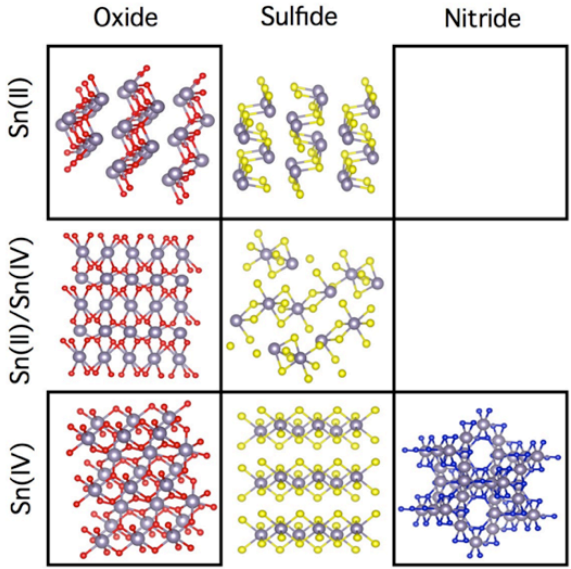

Fig. 1 Known and missing binary compounds of Sn. Whereas both Sn(II) and Sn(IV) as well as mixed Sn(II)/Sn(IV) oxides and sulfides have been reported, only the Sn(IV) nitride is known. The Sn(II) containing binary nitrides have not been reported in databases or literature.

Consider, for example, the useful and diverse chemistry of binary tin compounds. The oxides ($SnO_2$, $SnO$),[11,12] sulfides ($SnS$, $SnS_2$, $Sn_2S_3$)[13,14], and fluorides ($SnF_2$, $SnF_4$)[15,16] have been used for a wide range applications including transparent electronics[17] and batteries,[18] photocatalysis and antibacterial coatings,[19] photovoltaic absorbers[20] and contacts[21], as well as in dentistry[22] and periodontry.[23] This diversity is partially enabled by the propensity of tin to adopt three oxidation states: the metallic Sn(0), as well as the oxidized Sn(II) and Sn(IV). Tin nitrides are an interesting counter-example to the other Sn-based binary materials based on oxygen, sulfur or fluorine, since they are much less known (Fig. 1). The fact that this binary Sn-N materials family has missing members is somewhat surprising, given the Earth-abundance of constituent elements and more than 100 of years of extensive and systematic research in solid-state chemistry.

The search indicates that only the binary Sn(IV) material $Sn_3N_4$[24] is present in the ICSD crystallographic database.[25] A nitrogen-poor amorphous analogue to $Sn_3N_4$ has been also reported in literature.[26] As shown in Figure 1, binary crystalline compounds containing Sn(II), such as $Sn_3N_2$ (the pure Sn-II nitride) or SnN (one of the possible mixed Sn-II / Sn-IV) nitrides) are unknown so far. There are multiple possible structures for SnN (Sn IV/II) and $Sn_3N_2$ (Sn II) available in computational databases,[27,28] but they have yet to be experimentally realized. A record for the Sn IV/II nitride material (SnN) exists in the experimental ICDD[29] database, but the inspection of XRD patterns indicates it is actually the known $Sn_3N_4$ compound. We also note that up to date only one ternary Sn(II) nitride NaSnN[30] and one ternary Sn(IV) nitride $ZnSnN_2$[31,32] have been reported, so tin nitrides is a nearly virgin field of the solid state chemistry.

Another interesting aspect of the Sn-N material space is the metastability. The Sn(IV) nitride ($Sn_3N_4$) has a positive heat of formation with respect to the elements ($\Delta H_F$ = 1.6 eV/f.u),[33] meaning it thermodynamically favors decomposition into metallic tin solid and molecular nitrogen gas. Therefore any synthetic technique designed to produce SnN or $Sn_3N_2$ requires sufficiently reactive nitrogen and tin precursors to counteract thermodynamics. Furthermore, these precursors should be interacting at sufficiently reducing conditions to avoid the known $Sn_3N_4$, while maintaining quite oxidizing conditions to prevent complete reduction to Sn metal. All of this, including the very small binding energy difference between Sn 5s and Sn 5p electrons, and hence the required precise control of the oxidation state, makes *a priori* identification of the necessary Sn-N processing parameters difficult.

One way to locate an unknown processing window is to explore a wide range of processing space via combinatorial sputtering with spatial gradients in composition and other growth conditions.[34] Subsequent spatially-resolved characterization and high-throughput data processing elucidate composition, structure and properties of the films produced at each location.[35] Our previous work on films with temperature gradient[36,37] and target-substrate distance gradient[33,38] indicates that control of the cation oxidation state can be obtained by changing these variables in a binary materials system. Another way to address the problem of the expected but missing materials is by performing high-throughput theoretical structure prediction, but for non-ground state structures. Over the past several years, a number of ground sate structure prediction methods have been developed, including genetic algorithms,[39] data mining,[40] structure prototyping approaches,[41] and random/constrained structure samplers.[42] These first principles thermochemistry methods have led to several success stories,[5] for example the discovery of a large number of *hitherto* unknown ABX materials,[6] including those with Half-Heusler structures.[43]

The challenge of successfully utilizing the combinatorial thin film synthesis approach is that the crystal structure of the resulting samples may be difficult to determine. This is because the traditional approaches to solving the "inverse problem" of diffraction are hard to apply to materials in thin film form. The inverse diffraction problem is the task of calculating the crystal structure from its diffraction pattern, which cannot be solved by inductive methods. Therefore, it is traditionally solved deductively by (1) constraining the wide range of possible crystal structures to just a few candidates based on unit cell symmetry, and (2) subsequent direct calculation of the expected diffraction response of these few candidates.[44] Whereas this approach works well for single crystals or powder samples, in the case of thin films it may be complicated by broad diffraction peaks due to small grain size, and missing peaks due to preferential orientation. Hence a larger number of the crystal structure candidates needs to be considered theoretically, and additional experimental constraints need to be placed on possible candidates.

In this manuscript, we report on synthesis, properties, and possible crystal structure of a crystalline tin nitride material having a composition close to stoichiometric SnN, with a possible slight N-deficiency ($SnN_{1-\delta}$). Combinatorial reactive sputtering identifies that this SnN-like material is formed in the intermediate range of substrate temperatures (200-400 °C). In an attempt to determine the crystal structure of SnN, more than 6000 candidate structures were screened through three types of theoretical search methods employing density functional theory (DFT) calculations. These candidates were down-selected based on their calculated energy and on comparison to the experimentally measured long-range order (x-ray diffraction). The five most likley theoretical candidates were compared to the experimental sample by measuring a variety of local structure characteristics and physical properties. The results of this process suggest that the most likely structure of the new nitride is related to delafossite (space group 166), with stereo-active Sn(II) lone pair distortions and anion vacancies lowering the symmetry to space group 2. Interestingly, the SG2 candidate is 90 meV/atom above the lowest energy structure of the SnN. This result is rationalized by the quantitative structural similarity of the distorted SnN delafossite to the chemically related $Sn_3N_4$ spinel phase, providing new scientific insight into selection criteria for metastable polymorphs. In addition to the discovery of the novel binary nitride phase, another advance reported herein is the combined experimental/theoretical process for thin film based materials discovery and structure determination in non-equilibrium phase space.

**II. Methods**

Thin films of tin nitrides were grown on glass and silicon substrates by high-throughput combinatorial reactive sputtering of metallic tin targets in an argon and nitrogen atmosphere. Substrate temperature ($T_S$) and target-substrate distance ($d_{TS}$) were changed as orthogonal gradients, such that each position on the sample experienced different growth conditions. The resulting thin films were characterized by spatially-resolved X-ray diffraction (XRD) to determine the phase(s) present. After the phase space region of interest was identified, tin nitride films were prepared without growth-parameter gradients to obtain larger amount of material. Further local structure characterization was undertaken using Rutherford backscattering (RBS), Raman spectroscopy, electron microscopy, electron diffraction, X-ray absorption spectroscopy (XAS), and X-ray photoelectron spectroscopy (XPS). The tin nitride properties were measured using optical spectroscopy (infrared through ultraviolet) and Hall effect measurements. More information about our high-throughput combinatorial approach has been previously published elsewhere.[45,46] Additional details for all the combinatorial and single-point synthesis and characterization techniques are available in the Supplementary Information.

First-principles calculations were undertaken to identify candidate structures for SnN. Three separate routes were attempted with increasing sophistication and computational cost. The first approach involves using structure prototypes,[4,41] where SnN is assumed to take the form $ABX_2$ and A=Sn, B=Sn, and X=N. The second approach was Global Space Group Optimization, or GSGO,[47] which is a genetic algorithm for structure prediction. A third approach based on electronically and ionically biased random structure searching (RSS) was employed to elucidate possible non-equilibrium phases by searching a broader region of phase. The 4,000 of the SnN starting configurations of up to 20 formula units were generated using USPEX software[48], and 2,000 were generated using random superlattices (RSL) implemented in a recently developed polymorph sampling technique[49]. These structures were then geometrically relaxed using VASP software.[50] In the electronically biased RSS, an attractive or repulsive non-local external potential (NLEP)[51] was applied to the Sn 5s orbitals to influence them towards a 2+ or 4+ electronic state during the preliminary structural relaxation calculations. The ionically biased RSS uses RSL to preferentially search for cation-anion coordinated configurations that are expected to have large basins of attraction in configuration space, and hence an increased frequency of occurrence in random sampling and a larger probability to be realized during the synthesis experiments. Additional details of the theoretical methods are provided in Supplementary Information.

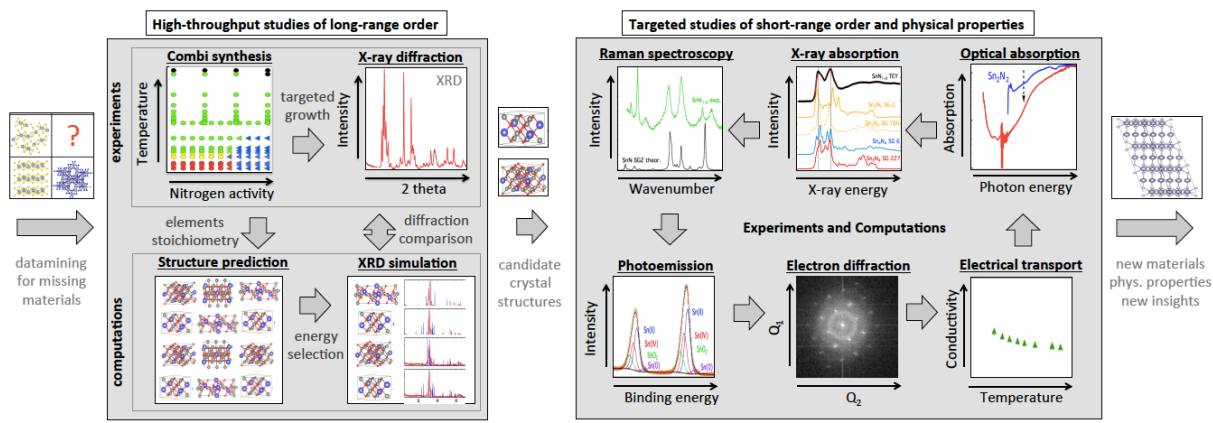

Fig. 2 Combined experimental/theoretical approach to identification of the crystal structure of new materials synthesized in thin film form. Compared to the traditional approach to solving the inverse problem of diffraction for crystals and powders, the number of initially considered structure candidates is much larger, and the solution is constrained by local structure measurements and physical property characterization.

The proposed combined theoretical and experimental process of solving the crystal structure of thin films is illustrated in Figure 2. In comparison with the traditional inverse diffraction problem solving process for crystals and powders, the main difference is the larger number of the synthesized samples and calculated structures (left), and an additional step of constraining the candidates by local structure characterization and physical property measurements (right). This schematic illustration (Fig. 2) also provides an outline for the remainder of this paper. First, the results of high-throughput experimental and theoretical studies are described, providing the sample stoichiometry and diffraction, as well as five most likely structure candidates, none of them exactly matching the measurement results. Next, four detailed characterization and calculation techniques are employed to evaluate the likelihood of each of these theoretical structures to be related to the experimentally synthesized material, pointing to the distorted SG2 version of the SG166 delafossite candidate structure. The optical absorption and electrical conductivity of the new SnN material are also reported. Finally, the results of these investigations are discussed in solid state chemistry terms, by quantifying the structural similarity between the proposed delafossite-related SnN and spinel $Sn_3N_4$ materials.

### III. High-throughput studies
#### A. Experiments

Experimentally, we found that changing both target-substrate distance and substrate temperature had an effect on the growth of the tin nitrides (Fig. 3). Substrate temperature ($T_S$) had the most dramatic effect on the phase of the resulting films, with spinel tin nitride ($Sn_3N_4$) observed at ambient temperature.[33] Increased substrate temperature resulted in the introduction of unassigned diffraction peaks with a decrease of intensity of $Sn_3N_4$ reflections, and a complete elimination of $Sn_3N_4$ reflections at $T_S = 120°C$. The highest intensity of the unassigned peaks was observed around $T_S = 350°C$. These reflections did not match any known tin or tin nitride phase, or any known compound containing other possible elements (Sn, N, O, H). Interestingly, at $T_S >$ 450°C, $Sn_3N_4$ is again observed, in this case together with Sn metal ($\beta$-polymorph). This suggests that the tin-nitrogen bonds in $Sn_3N_4$ are more thermally stable than those in the new material; in other words the newly observed material should be even less thermodynamically stable than $Sn_3N_4$.

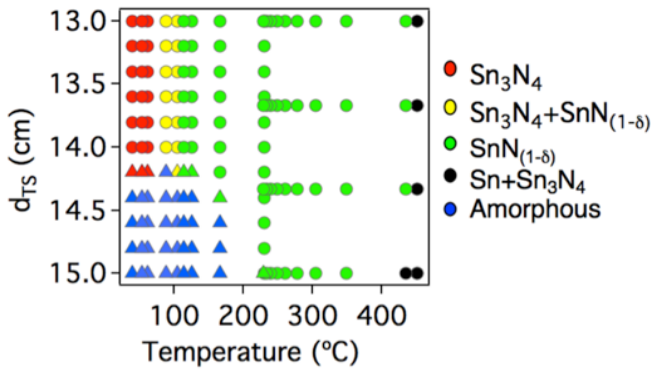

Fig. 3 Combinatorial thin films synthesis growth space (phase diagram) of Sn-N films obtained at different substrate temperatures. Both the substrate temperature and the target-substrate distance had an effect on the resulting phase and preferred orientation (Figure S10).

The phase assignments were made by XRD measurements (see Supplementary Fig. S1a for more details).

When target-substrate distance was changed at a given substrate temperature, differences in crystallographic texture were observed in the two-dimensional XRD detector images (Figure S10). Small target-substrate distances ($d_{TS}$ = 13-14 cm) generally produced films with increased long-range order compared to films grown at larger target-substrate distances ($d_{TS}$ = 14-15 cm). Such preferential orientation effects often complicate the crystal structure determination of thin film materials. The XRD patterns as a function of $d_{TS}$ are shown in Supplementary Figure S1a. Interestingly, despite the strong crystallographic texturing, peak broadening indicative of weak long-range order or small grain size was observed in all films.

The stoichiometry of this new tin nitride material synthesized at intermediate temperature was subsequently measured by the RBS to be between SnN and $Sn_9N_8$, suggesting the mixed-valent Sn-IV / Sn-II nitride. The resulting RBS spectra are shown in the Figure 4. The range of observed Sn-N stoichiometries is presented in the inset of Fig.4, in comparison with measured $Sn_3N_4$ composition (Sn-IV) and expected $Sn_3N_2$ composition (Sn-II). Differences between individual samples, oxygen contamination, and instrumental uncertainty could have contributed to the range of measured composition of the Sn-N samples. No correlation between measured stoichiometries and variations in XRD patterns was observed. Based on this RBS analysis, we refer to the new synthesized material as $SnN_{1-\delta}$ ($0 < \delta < 0.2$) for the remainder of the manuscript.

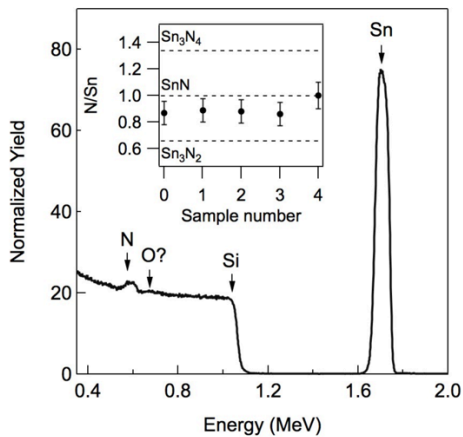

Fig. 4 Rutherford backscattering spectrum, showing a strong Sn peak and a weaker N hump on top of the Si step edge. Inset: the Sn-N sample composition statistics, indicating the thin film composition close to SnN. The $SnN_{1-\delta}$ sample is more nitrogen-poor compared to the measured $Sn_3N_4$ reference sample, and more nitrogen-rich compared to the expected $Sn_3N_2$ stoichiometry.

To address the possibility that $SnN_{1-\delta}$ may be more than one new phase, a film was grown isothermally at 350°C and subjected to a high-throughput anneal by applying a post-growth temperature gradient across the substrate in an inert atmosphere. Subsequent spatially resolved XRD analysis of the film revealed the structural changes at each temperature. The XRD patterns as a function of the annealing temperature are shown in Figure S1b. The sample underwent a decomposition near 450°C, into Sn and $Sn_3N_4$. The abrupt disappearance of all $SnN_{1-\delta}$ diffraction at a single temperature indicates that the diffraction peaks all belong to a single phase or that all phases have similar kinetic stability. Such disproportionation is also observed in the tin oxide system, with the metastable intermediate oxide $Sn_3O_4$ transforming to the thermodynamically stable Sn metal and $SnO_2$ upon heating.[52]

To investigate the crystallinity of $SnN_{1-\delta}$ further, a powder sample was prepared by growing several thick films at one condition ($T_s$= 350°C and $d_{TS}$= 13 to 14 cm) and scraping the films from the substrates. XRD analysis of the resulting powder revealed additional peaks not detected in the as-grown films, confirming the preferential crystallographic orientation observed by the area detector (Figure S10). The powder diffraction pattern is shown in Figure 5a, with a cluster of peaks in 2Θ=30°-40° and 60°-70° degree range. Close inspection of these clusters reveals that their individual peaks do not overlap with $Sn_3N_4$ or Sn reflections. Further, the presence of the Sn and $Sn_3N_4$ secondary phases can be ruled out based on the absence of their expected strong peaks in the 40°-60° range, as indicated by vertical lines in Fig. 5. We also note that the $SnN_{1-\delta}$ XRD pattern contains a large number of peaks pointing to low symmetry, multiple SnN phases, or both. Also, the absence of peaks below 2Θ~30° indicates a small unit cell. The summary of the 2Θ values of the measured XRD peaks and the associated calculated $d$-spacings is provided in Table S2.

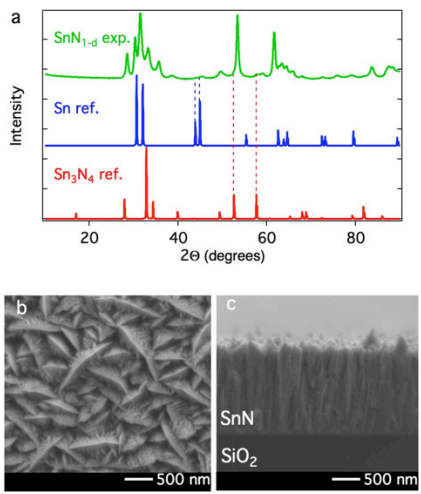

Fig. 5 (a) XRD patterns of $SnN_{1-\delta}$ powders scraped from a large number of thick $SnN_{1-\delta}$ films, indicating a small unit cell with low symmetry. The absence of Sn and $Sn_3N_4$ impurity phases is indicated by vertical dashed lines from the corresponding reference patterns at higher angles. The broad peaks can be explained by small grains observed using scanning electron microscopy (SEM) as shown in (b) top view and (c) cross-sectional images.

The relatively broad peaks in XRD suggest a small coherence length, and makes traditional Rietveld refinement with these samples impossible. Monochromatic synchrotron radiation with high intensity did not improve the width of the XRD peaks, supporting the small grain hypothesis, rather than the detector- or source-limited resolution of the lab XRD setup. Indeed relatively small microstructural features with 500 nm size can be observed in top-down the SEM images (Fig. 5b), and even smaller <100 nm microstructure is visible in cross-sectional SEM (Fig. 5c). Thus, with the knowledge the composition from RBS and the likely single phase from thermal decomposition, we thus turned to high throughput theory to generate a large set of SnN candidate structures.

### B. Theory

To identify the unknown SnN structure we started by looking for the low energy ones using structure prototyping search (SPS) and genetic algorithm (GSGO) methods. The choice of these methods was motivated by both their tractable computational costs, as well as our previous familiarity with these techniques for structure prediction of previously unreported thermodynamically stable "missing" materials.[6,41,43,4] The SPS computations using $ABX_2$ compounds (A=Sn, B=Sn, and X=N) identified two structures shown in Figure 6a, with energies comparable to other SnN candidates and $Sn_3N_4$ (see Table S1 in Supplementary Information). Even though the simulation of the XRD patterns of these two structures did not find the exact match to the experimental $SnN_{1-\delta}$ pattern, the lowest energy $CuBiS_2$ (SG62) structure could not be conclusively ruled out, so it was retained for further theoretical characterization. In contrast to the SPS technique, the GSGO method identified the low-energy SnN structure to be the decomposed form of Sn- and N clusters, consistent with the expected positive formation enthalpy for SnN compounds with respect to Sn metal and $N_2$ molecule. This shows that genetic algorithms in their standard forms are more suitable for structure prediction of thermodynamically stable materials than the metastable materials discussed here.

The inconclusive findings from the GSGO and unsatisfactory XRD match from the SPS methods motivated us to consider other techniques to identify the candidate SnN polymorph structures. The condition of low formation energy of the GSGO approaches or the restriction to known crystal chemistries of SPS methods impose adverse constraints on the viable phase space used to identify the structures of new metastable materials. Even for the GSGO-like methods aimed to enhance the sampling of metastable configurations, such as biased metadynamics[53] or antiseeding,[48] the abundance of possible low energy configurations corresponding to metallic tin and molecular nitrogen pose a challenge to defining the optimal calculation parameters. Thus, alternative structure predictions methods based on the designed for finding metastable materials, such as the random structure search (RSS) methods, were needed.

In order to overcome the GSGO and SPS limitations, we used electronically- and configurationally biased random structure search (RSS) to identify candidate metastable SnN polymorph structures, as described in the method section. Taking into account the mixed Sn(II)/Sn(IV) character of the SnN material, we performed an electronically biased RSS of over 4,000 randomly generated SnN, $Sn_2N_2$, $Sn_4N_4$, and $Sn_8N_8$ unit cell structures[48] with electronic bias on unique Sn atoms towards a 2+ or 4+ electronic state.[51] Additionally, we generated 2,000 structures using the ionically biased RSS technique, also referred to as random super-lattice (RSL) polymorph sampling.[49] To preferentially avoid metallic tin and molecular $N_2$ configurations in all 6000 of calculations, we imposed constraints on the interatomic distances of Sn-Sn and N-N to be > 2.0 Å in the randomly generated structures. The results of both RSS calculations were then filtered to remove identical structures or those that contained phase-separated Sn-metal / N-dimer regions. All of the electronic- and ionic bias constraints were lifted prior to the completion of the structural optimization procedure.

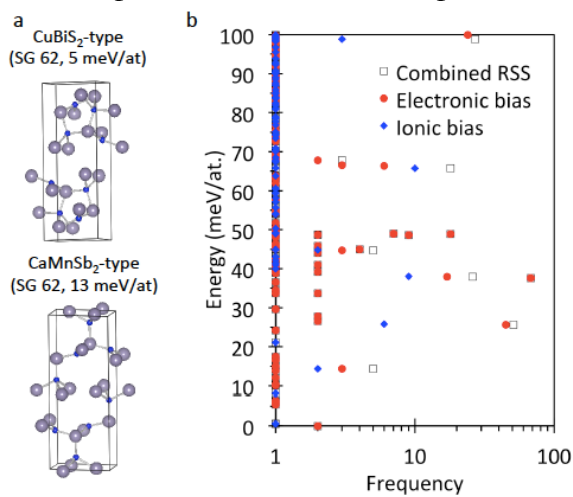

Fig. 6 The results of the theoretical SnN structure search methods. (a) The two low energy structures without the N-N dimers, identified from $ABX_2$ structure prototyping search method. (b) Energy distribution of SnN structures (excluding identical and phase-separated compounds) from electronically and configurationally biased random structure sampling techniques.

The energy distribution of the resulting geometrically optimized structures in the 100 meV energy range is depicted in Figure 6b, including the results from both RSS methods, and excluding identical and phase-separated compounds. The wider energy range is shown in Figure S9. The ionically biased RSS method resulted in a tighter distribution of candidate structures grouped within lowest 400 meV/atom energy range sampled. This indicates a much more selective sampling of low energy polymorphs as compared to the electronically biased sampling; however both methods did identify several low-energy high-probability structures. In addition, the electronically biased RSS method also identified many structures that the ionically biased RSS method did not find, but only with a low frequency of occurrence and expected low basin of attraction in configuration space. This comparison further supports the statistically motivated premise of the RSL polymorph sampling approach.[49] As intended, the electronic biasing of the RSS resulted in only a small fraction of the predicted low energy structures being composed of either metallic Sn or molecular nitrogen phases. Based on these results (Fig. 6b), we identified the lowest energy SnN polymorph ($Sn_4N_4$-SG12) from all of our search techniques. This structure was used as the lowest energy reference point throughout the remainder of this paper.

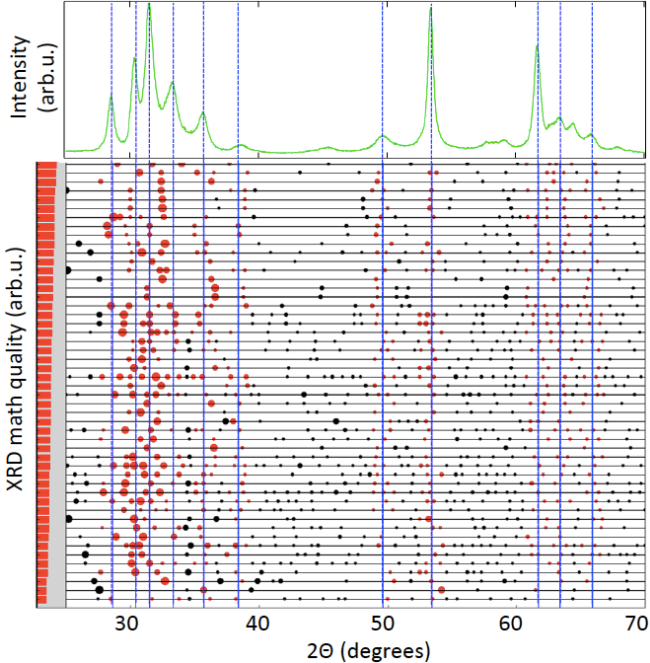

Fig. 7: The list for fifty low-energy SnN structure candidates ranked by their match to the experimentally measured $SnN_{1-\delta}$ XRD pattern using an automated search-match algorithm. The space groups of the top five structural candidates are 6, 12, 25, 62, and 166, and their XRD patterns are shown in Figure S8

To further narrow down the list of likely SnN crystal structures from 100s (Fig. 6b) to just a few (Fig. 7), we compared the simulated XRD patterns to the experimentally obtained powder pattern. This was done for the 50 lowest-energy unique structures generated using a high-throughput peak-matching algorithm (see Supplementary Information for more details). The algorithm provided a rank list of the matches, thereby identifying the most likely crystal structures that subsequently can be used for higher-level computations. Although no simulated XRD pattern was found to be in exact agreement with the experimentally measured pattern, five most likely candidates emerged from the sorting procedure. These candidates had space groups 6, 12, 25, 62, and 166, and all had energies within 100 meV/at. of the SG12 lowest energy polymorph. Their XRD patterns are shown in Figure S8 along with the experimental pattern. The XRD comparison indicates that the SG6, SG12 and SG62 candidates are less likely than the SG25 and SG166 candidates, since the former have XRD peaks in low-angle range, which is not observed in the experimental XRD pattern. In addition to these five SnN candidate structures, we chose to include $Sn_3N_4$ (space group 227) in our higher-level calculations. With this information in hand, we further evaluated the 5 candidate SnN structures by simulating their local structure and physical properties, and compared them to the corresponding experimentally measured properties, as described next.

## IV. Detailed studies
### A. *Raman scattering*

To elucidate the local atomic structure of the Sn-N samples, we turned to Raman scattering – a useful fingerprinting technique for assessing local structure in solids that is complementary to long-range XRD characterization. The more local Raman measurements are especially useful for thin film samples with small grains or poor crystallinity. However, the Raman also has certain disadvantages compared to the XRD: whereas XRD has a massive set ($>10^5$) of reference patterns that can be obtained from structural databases such as ICSD[25] and ICDD[29], the number of available Raman reference patterns lags behind and is scattered through the literature. In addition, depending on the local symmetry and free carrier concentration, some materials may show no Raman response. All these factors make routine fingerprinting of the newly synthesized materials using Raman scattering operationally challenging.

The results of the Raman scattering measurements on the $SnN_{1-\delta}$ samples are shown in Figure 8. Two pairs of peaks in the 350-550 cm$^{-1}$ and 600-800 cm$^{-1}$ were observed, in addition to several other peaks at lower wavenumbers (100-200 cm$^{-1}$). The corresponding first-principles simulations on the initial set of five SnN candidates and $Sn_3N_4$ reference patterns did not show any obvious matches to experimental data (Fig. S5). Thus, based on this comparison all 5 candidate SnN structures as well as $Sn_3N_4$ contamination in the $SnN_{1-\delta}$ samples could be ruled out. However, we noticed that one of the structural candidates (SG166) had two strong peaks in the 400 cm$^{-1}$ and 600 cm$^{-1}$ regions close to the experimentally observed peak doublets, suggesting that it may be related to the $SnN_{1-\delta}$ sample.

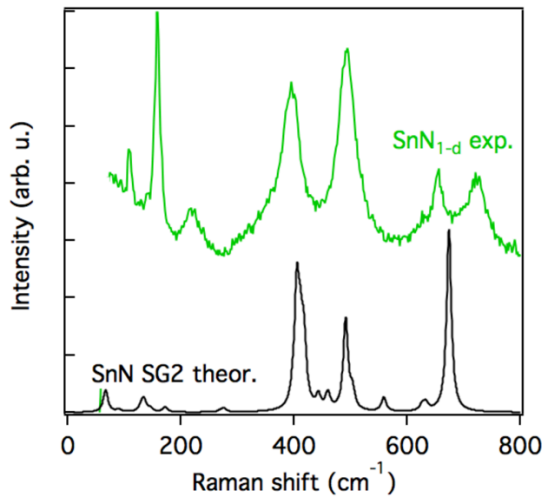

Figure 8. Experimentally measured Raman spectra of the $SnN_{1-\delta}$ samples, in comparison with the theoretically simulated response of the SnN SG2 structure, a distorted version of the SnN SG166 candidate. The structural relaxation of the SG166 supercell resulted in the splitting of the Raman modes, leading to better qualitative agreement of the SG2 structure with the experimental measurements.

To expand on this Raman observation, the SG166 candidate structure was found to be mechanically unstable, as evidenced by negative frequencies in phonon calculation results (Figure S6). The subsequent supercell calculations with further relaxation revealed the same imaginary phonon mode and led to a lower energy relaxed structure. These relaxations lowered the energy by 5 meV/atom and resulted into breaking the symmetry of the SG166 candidate down to the SG2 (Figure S7). In turn this led to splitting of the calculated Raman peaks, and hence a qualitatively better match with the experimentally measured spectra. We note that the SG2 theoretical peaks were still shifted to the lower wavenumbers compared to the experimental measurements, due to overestimation of the lattice constant in GGA DFT. However, the addition of the Van der Waals correction improved the quantitative agreement (Fig.8), in particular for the doublet at lower wavenumbers. This symmetry-broken SG2 structure was used in subsequent theoretical calculations such as x-ray absorption (XAS) reference spectra discussed next. It should be stressed that the symmetry breaking can happen in several directions and since the energy minima are very close - hence the actual SnN structure might show some degree of disorder, and yet lower energy due to entropic stabilization. This can also lead to better agreement for the doublet at higher wavenumbers (Fig. 8).

### B. X-ray absorption spectroscopy (XAS)

To further characterize the local structure of the $SnN_{1-\delta}$ samples, we used x-ray absorption spectroscopy (XAS). In the XAS measurement, the samples are irradiated with x-rays of different energies close to the N 1$s$ absorption edge, and the resulting photoelectrons from de-excitation are measured. Similar to the x-ray photoelectron spectroscopy (XPS), the XAS measured from the total electron yield (TEY) is a relatively surface sensitive measurement: a mean free path of the photoexcited electrons on the order of tens of nanometers. The strength of the XAS method compared to the lab-based XPS is that the final state selection rules allow for a straightforward calculation of the spectra from a given structure.

The results of the $SnN_{1-\delta}$ XAS measurements were compared with the theoretical simulation of the XAS spectra for different SnN structure candidates predicted from theoretical calculations, as well as $Sn_3N_4$, to determine which theoretically predicted structures are most likely present in the $SnN_{1-\delta}$ sample. The results of this comparison are provided in Fig. 9 and Fig. S2. Experimentally, $SnN_{1-\delta}$ shows two strong peaks separated by ~7 eV close to N 1s x-ray absorption edge. This double-peak x-ray absorption feature is consistent with the SG6, SG2 and SG166 candidate structures of SnN, and with SG 122 spinel $Sn_3N_4$ (Fig. 9). The other structure candidates (SG25, SG15, SG62) have only the lead XAS peak (Supplementary Fig. S2), and thus can be present in the $SnN_{1-\delta}$ sample, but cannot be its single phase.

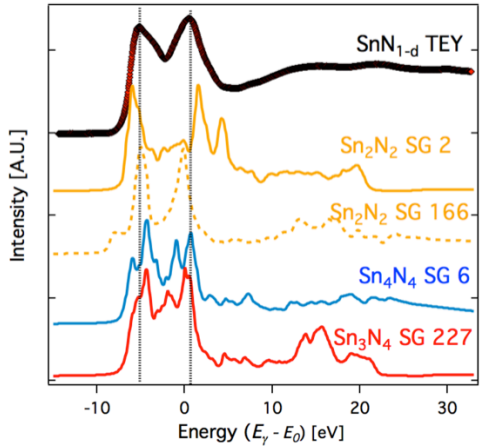

Figure 9: Nitrogen 1$s$ x-ray absorption spectroscopy of the $SnN_{1-\delta}$ samples, in comparison with the theoretically calculated response of the SnN in different theoretically predicted structures. The SG2, SG166 and SG6 structures (and possibly $Sn_3N_4$) are more likely, whereas SG25, SG12 and SG62 are less likely (see Supplementary Fig. S2)

### D. X-ray Photoelectron Spectroscopy (XPS)

The XPS was undertaken to elucidate the elemental composition and the bonding environments of the tin and nitrogen species in the $SnN_{1-\delta}$ samples. The major advantage of XPS is that it enables quantification of these materials characteristics, and thus facilitates comparisons between the material of interest (here $SnN_{1-\delta}$) and the related materials (here $Sn_3N_4$) or the modeled crystal structures (here SnN). One potential disadvantage is that the lab-based XPS with monochromatic source is a rather surface sensitive technique with a few nm information depth, so the presence of native oxide at the surface complicates analysis. Whereas XPS with subsequent peak fitting is a very common technique to carefully analyze valence states of elements in oxides, such thorough analysis of nitrides is less commonly reported in literature, and is presented here.

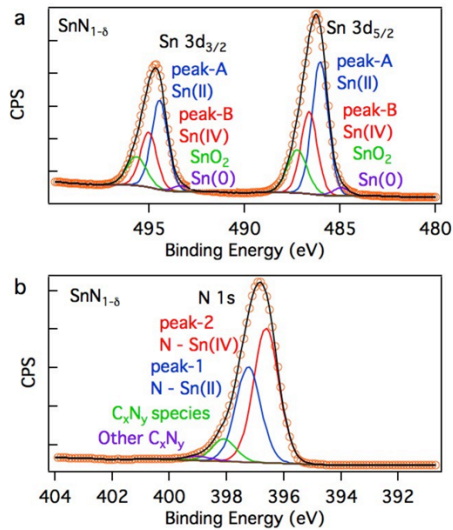

Figure 10. High-resolution XPS (a) Sn 3d and (b) N 1s spectra for SnN$_{1-\delta}$ sample and the corresponding peak fitting results. The peak fitting suggests the presence of two valence states of tin, Sn(IV) and Sn(II), and two distinct N components, one associated with Sn(IV) and another with Sn(II). Together, these results indicate mixed-valent SnN structure with two distinct N atom sites.

The high-resolution Sn 3d spectra shown in Fig.10a suggests two major nitride components, one associated with Sn(II) indicated by peak-A at 486.0 eV, and another with Sn(IV) indicated by peak-B at 486.6eV. Analysis of the N 1s spectra (Fig. 10b) also shows two major components, indicative of two different chemical/structural environments. The first component located at higher binding energy (peak-1 at 397.3 eV) is assigned to nitrogen bonded predominantly to lower valency Sn(II), while the second component located at lower binding energy (peak-2 at 396.6 eV) is assigned to nitrogen bonded more closely to higher valency tin Sn(IV). These peak assignments are consistent with literature[54,55,56,57] and similar trends were observed for other metal nitrides.[58,59,60]

Table I. Measured concentrations of Sn(II) and Sn(IV) in SnN$_{1-\delta}$ and Sn$_3$N$_4$ samples, as well as predicted atomic concentration of nitrogen associated with Sn(II) and Sn(IV) calculated from a N:Sn ratios of 3:2 and 4:3 respectively. These are compared to the XPS determined atomic concentration of nitrogen associated with Sn(II) and Sn(IV) from peak fits of the N 1s spectra

| Sample composition | SnN$_{1-\delta}$ | Sn$_3$N$_4$ |
|---|---|---|
| Measured total Sn | 29.8±0.2 | 30.3±0.9 |
| Fitted peak-A Sn-II | 13.3±1.0 | 13.2±1.0 |
| Fitted peak-B Sn-IV | 9.3±0.4 | 12.3±0.4 |
| Measured total N | 24.1±0.2 | 28.5±1.9 |
| Fitted peak-1 N(Sn-II) | 9.0±0.1 | 8.8±0.8 |
| Fitted peak-2 N(Sn-IV) | 12.8±0.1 | 14.7±1.0 |
| Calculated N(Sn-II) | 8.9 | 8.8 |
| Calculated N(Sn-IV) | 12.4 | 16.4 |

To confirm the peak assignments, the atomic percent of nitrogen determined from each of the Sn peaks associated with tin nitride were calculated assuming 4:3 N:Sn and 2:3 N:Sn ratios. These calculated values were found to align well with values determined from the measured N 1s spectra (Table I). This experimental observation suggests that there are two unique N sites in this mixed-valent tin nitride, an important insight for the candidate SnN structures. Indeed, all five theoretically proposed candidate structures (including SG2 structure, the distorted SG166 candidate) have two unique nitrogen sites, which is consistent with this experimental observation.

Comparing the SnN$_{1-\delta}$ XPS results with those for the Sn$_3$N$_4$ reference sample (Supplementary Fig. S3), we note that the SnN$_{1-\delta}$ shows higher amounts of Sn(II) components relative to the Sn$_3$N$_4$, confirming major differences in the composition between the two films. Nevertheless some degree of surface contamination is present in both samples. This is indicated by the relatively weak Sn metal-like components, small amounts of $SnO_2$ in Sn 3d spectra, and the nitrogen associated with carbon (C-N functionalities) observed in N 1s spectra.

### D. Transmission electron microscopy (TEM)

In order to gain additional insight on crystallographic structure, microstructure, and morphology of the $SnN_{1-\delta}$ samples, we performed transmission electron microscopy (TEM) and selected area electron diffraction (SAED) on cross-sectional samples of the material. The advantage of these methods over conventional XRD is that they provide spatially-resolved structural measurements, which helps elucidate phase-purity and crystallinity of the sample at smaller length scales. A potential drawback is that the material may be damaged during sample preparation via focused-ion beam milling, and that TEM/SAED preparation and analysis can be quite time-consuming procedures.

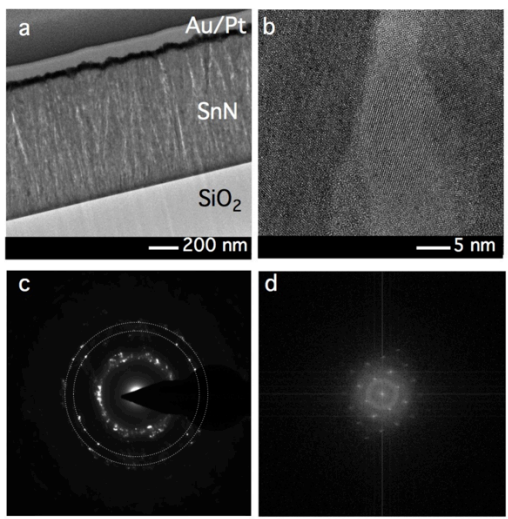

Fig. 11 (a) Cross-sectional TEM image of the tin nitride sample at low magnification. (b) High-resolution TEM image of the sample showing three separate grains/areas with different contrast. (c) Representative SAED pattern from the $SnN_{1-\delta}$ area in (a), showing diffraction spots on two rings with small *d*-spacing (indicated by dashed circles), and several rings with large *d*-spacing, consistent with XRD in Fig. 5a. (d) Representative FFT of the single-contrast shown area in (b), showing that small regions of the material are single crystals.

The TEM imaging shown in Figure 11 (a-b) indicates that the $SnN_{1-\delta}$ layer is polycrystalline, with ~100-1000 nm sized columnar regions that consist of ~10 nm sized crystalline particles surrounded by more disordered material. These TEM observations are consistent with the SEM results (Fig. 5 b-c). Similar small-grain microstructures have been observed in other novel nitride-[33,38] or oxide[61,62] materials in their initial stages of development. In order to analyze small areas of the sample that were inaccessible via SAED, we performed fast Fourier transforms (FFT) of several single particles from the high-resolution TEM images (represented in Fig. 11b). The results of this analysis (Fig. 11d) indicate that crystalline areas of the $SnN_{1-\delta}$ samples are a single phase. Furthermore, the SAED/FFT analysis suggests that SnN crystal structure has a small unit cell with low symmetry, consistent with conclusions from the prior XRD analysis (Fig. 5a). Specifically, one possibility is that the unit cell parameters of $a$ = 4.22, $b$ = 4.12, $c$ = 3.14, $\alpha$ = 83.5, $\beta$ = 91.2, and $\gamma$ = 92.1, which would reproduce many of the experimentally observed *d*-spacings presented in Table S2. However, this analysis is limited by the identification of only four unique imaged zone axes from the HRTEM images (Fig. 11).

### *E. Optoelectronic properties*

To further constrain the possible SnN candidate structures, and to evaluate the potential future applications of $SnN_{1-\delta}$ thin films, we measured and calculated its optoelectronic properties. The experimental optical and electrical properties of the $SnN_{1-\delta}$ samples suggest a degenerate *n*-type semiconductor with a band gap between 1 and 2 eV. The results of these characterizations are shown in Figure 12. The $SnN_{1-\delta}$ film has a shallow absorption onset above 1 eV with an inflection close to 2 eV. At energies below 1 eV, the extinction also increased as the result of free-carrier absorption. This indicates that the actual band gap of this material may be smaller than the optical absorption onset, due to band filling effects (Burstein-Moss shift).

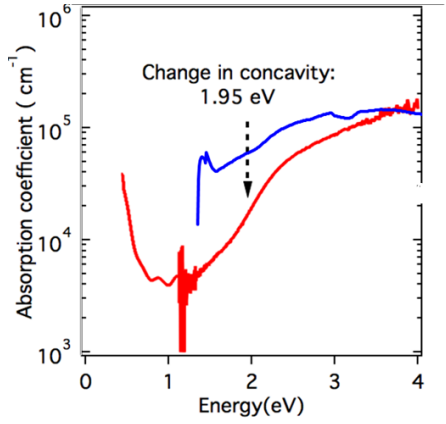

Fig. 12 Measured optical properties of $SnN_{1-\delta}$ (red), and the simulated adsorption (blue) of the delafossite-related SG2 SnN candidate structure.

Free carriers were also observed in the temperature-dependent Hall measurements (Fig. S4), where decreasing the temperature has almost no effect on the conductivity. Room temperature Hall effect measurements showed an *n*-type carrier concentration of $3\times10^{20}$ cm$^{-1}$ and a mobility of 2 cm$^2$/Vs. This suggests that the optoelectronic properties observed here originate from the crystalline grains rather than the low-order grain boundaries observed in TEM (Fig. 11), because previous measurements on amorphous tin nitride have shown a lower mobility by a factor of five.[38] All these optical and electrical experimental observations are consistent with the theoretical electronic structure calculations that determined that SnN with the distorted delafossite SG2 structure is a semimetal (Table II). The calculation results for the optical absorption spectra of the SnN in SG2 structure are shown in Fig. 12. In comparison with the experimental absorption spectra, similar slow onset can be seen, further supporting the SG2 as the most likely structural candidate for the SnN material.

## V. Discussion

The results of all the characterization and calculation techniques are summarized in Table II, along with the calculated energies of the candidate structures. It appears that among the 5 different possible SnN candidates identified by the theoretical calculations, the SG2 structure (distorted SG166 candidate) is the most likely one to describe the $SnN_{1-\delta}$ thin films. The proposed SG2 structure of SnN consists of sheets of octahedrally bonded Sn(IV) ions alternated with the linearly coordinates Sn(II) ions along the $c$-axis of the layered distorted delafossite-like structure (SG166). The distorted-SG166 candidate (SG2) is supported by Raman and XAS measurements, and it is also consistent with the XRD, XPS, SAED characterization. However, not all the $SnN_{1-\delta}$ thin film XRD peaks can be explained by the SG2 structure, indicating that either there are further structural distortions (see discussion in Raman section above), or that previously unreported secondary phases are present. The possibility that the $SnN_{1-\delta}$ thin films contain some amount of the $Sn_3N_4$ phase, can be ruled out by the theoretical and experimental XRD patterns (Fig. 5) and Raman spectra (Fig. 8 and Figure S5) of $Sn_3N_4$, which are inconsistent with the $SnN_{1-\delta}$ thin films measurement results. So given the overall close to Sn:N=1:1 stoichiometry of the samples (Fig. 4), the secondary phase (if any) is likely to also have the SnN composition, which can be one of the SnN polymorphs listed in Table II.

Table II A summary of theoretical results for top five SnN candidate structures based on the XRD peak search-match algorithm, in comparison with the experimental observations for the $SnN_{1-\delta}$ sample and theoretical calculations for $Sn_3N_4$. The most likely structure of the $SnN_{1-\delta}$ samples is the distorted SG2 SnN structure related to the delafossite SnN (SG166) that is 96 meV/at. higher in energy than the lowest energy SG12 structure

|  | $Sn_3N_4$ | SG12 | SG62 | SG6 | SG25 | SG166 | SG2 | expt. $SnN_{1-\delta}$ |
|---|---|---|---|---|---|---|---|---|
| $H_F$, meV | 0 | 0 | 5 | 6 | 55 | 96 | 90 | >$Sn_3N_4$+Sn |
| XRD | no | no | no | no | maybe | maybe | maybe | broad peaks |
| Raman | no | maybe | no | maybe | no | no | yes | two doublets |
| XAS | yes | maybe | maybe | yes | maybe | yes | yes | two peaks |
| SAED | no | yes | maybe | yes | maybe | maybe | yes | small u.c. |
| XPS | no | yes | yes | yes | yes | no | yes | two N peaks |
| Electrical | semicond. | semicond. | semicond. | semimet. | metal | semimet. | semimet. | degenerate |
| $E_g$, eV | 1.8 | 0.9 | 0.9 | 0.8 | 0.0 | 1.4 | 1.4 | 1.0 - 2.0 |
| Similarity | 1.00 | 0.43 | 0.49 | 0.57 | 0.53 | 0.75 | 0.78 | likely |

It is quite unexpected that SnN with the SG2 structure turned out to be the most likely candidate for the $SnN_{1-\delta}$ samples, since its energy is almost 0.1 eV/atom higher than the lowest-energy SG12 candidate (Table II). As show in Fig. 13a, both SG12 and SG2 SnN are also higher in energy than the $Sn_3N_4$-Sn ground state line on the pseudo convex hull, between the metallic Sn and atomic N present in the sputtering chamber during the synthesis process. Recall that even $Sn_3N_4$ is higher in energy than the Sn metal and $N_2$ molecule,[33] so overall all the compounds observed here in the Sn-N materials system are highly metastable.

One possible reason is that the SG2's parent SG166 structure gets stabilized by templated heterogeneous nucleation out of the structurally similar $Sn_3N_4$ spinel phase (Fig. 13b), which might be present at the initial stages of growth. To quantify the relationship between the $Sn_3N_4$ spinel and the different SnN polymorphs, we employ a structural similarity function[63] based on Voronoi decomposition[64] of a crystal structure into a set of substructures. A structural similarity value close to 1.0 indicates high geometric similarity between substructural polyhedra.

As shown in Table I, the similarity between $Sn_3N_4$ spinel and the SnN SG166 candidate is the highest of the considered candidate polymorphs, and it is even higher for the distorted SG2 structure. This high similarity originates from the presence of octahedrally coordinated Sn in both $Sn_3N_4$ spinel and SG166 delafossite (or SG2 distorted delafossite), which is not the case in the four other lower energy polymorphs. Crystals of a given chemistry tend to exhibit similar coordination environments,[65] suggesting that SG166/SG2 is preferred over the other lower-energy polymorphs for structural reasons, rather than purely energetic reasons. These results suggest that the structural similarity metric may complement traditional energetic considerations as a heuristic for identifying which metastable structures nature tends to form.

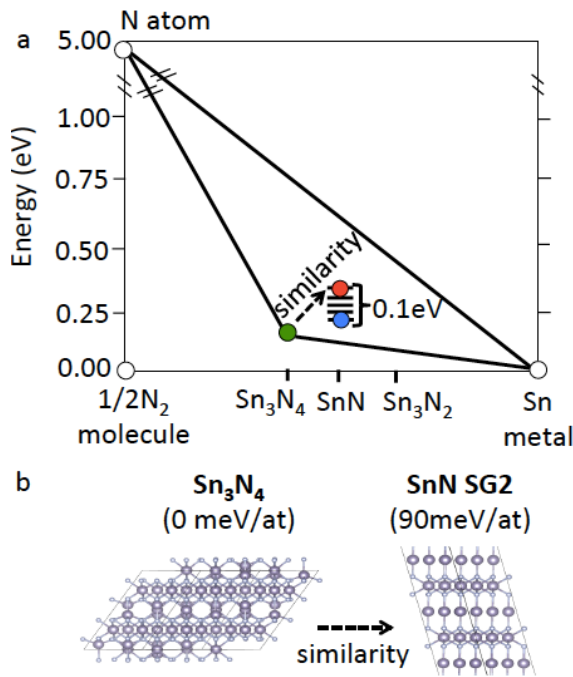

Fig. 13 (a) Schematic illustration of the Sn-N pseudo convex hull with respect to Sn metal and N atoms present during the sputtering process. It shows that the SnN SG2 structure (red circle) that is most similar to the related $Sn_3N_4$ (green circle) was experimentally realized, rather than the lowest energy SG12 SnN (blue circle). (b) The spinel and the delafossite crystal structures, showing the similar layers of octahedrally coordinated atoms, and supporting the structural similarity conclusion from the theoretical algorithms

The SG166 delafossite crystal structure related to SnN is well known primarily for supporting both optical transparency and *p*-type conductivity in the Cu-based ternary oxides such as $CuAlO_2$.[66] In addition, the oxide delafossites have been also reported for visible light photocatalysts,[67] ozone sensors,[68] and other applications (thermoelectrics, catalysts, and antibacterial coatings). Very recently, nitride delafossites such as $CuTaN_2$[69] and $CuNbN_2$[70] have been proposed as potential light absorber materials for photovoltaic solar cells and photoelectrochemical water splitting, and subsequently as thermoelectrics.[71] The SG2 SnN reported here is a new addition to the versatile family of the delafossites. Whereas its degenerate semiconducting properties and possible semimetal character may impede its future utility as a light absorber in optoelectronic devices, it still may be interesting for some of the other aforementioned applications.

## VI. Summary and Conclusions

In summary, this paper reports on the synthesis, characterization, and the attempts of structural identification of a novel earth-abundant semiconductor in the Sn-N family. The experiments indicate that the material is likely the nitrogen-deficient mixed-valent Sn(II/IV) nitride $SnN_{1-\delta}$. The SnN structure prediction methods, in comparison with the experimental results, suggest that the distorted space group 2 structure (SG2), related to the delafossite structure (SG166), is most likely. While we have made many efforts to determine the crystal structure of this material using a large number of experimental and theoretical methods, we must leave further structure refinement, including atomic position determination, to future work. Larger crystals and fewer defects should assist in crystal structure determination, and would also lead to better understanding of possible applications of this material. Other research groups are cordially invited to perform such studies.

To conclude, the discovery of a novel binary semiconductor composed of Earth-abundant elements and synthesis at mild conditions demonstrates that the periodic table still holds surprisingly simple compounds awaiting discovery. This realization calls for more such research in the underexplored materials chemistries like nitrides, which can be accessed by thin film synthesis approaches. The reported structural similarity comparison used to rationalize the SnN SG2 polymorph selection gives fresh insight into possible reasons for realization of the metastable materials, beyond the simple lowest energy criteria. The combined experimental/theoretical study presented in this work, including generation and ranking of a large number of structural candidates based on the long-range order, a variety of local structure characterization/calculation techniques, all demonstrate a new approach to determining the crystal structure of thin film materials.

**Acknowledgements**

This work is supported by the US Department of Energy, Office of Science, Basic Energy Sciences, under Contract no. DEAC36-08GO28308 to NREL as a part of the DOE Energy Frontier Research Center "Center for Next Generation of Materials by Design: Incorporating

Metastability." X-ray absorption measurements were performed at Stanford Synchrotron Radiation Lightsource at SLAC National Accelerator Laboratory, supported by the U.S. Department of Energy, Office of Science, Basic Energy Sciences under Contract No. DE-AC02-76SF00515. We would also like to thak Kevin H. Stone for attempting synchrotron XRD measuremetns of the $SnN_{1-\delta}$ samples at SLAC. The authors acknowledge the use of instruments at the Electron Imaging Center for NanoMachines supported by NIH (1S10RR23057) and the California NanoSystems Institute at UCLA, with raw data taken by Chilan Ngo, who thanks Suneel Kodambaka at UCLA for useful discussions, and gratefully acknowledges support from the NSF through Grant No. CMMI-1200547. The XRD peak search-match algorithm used in this work was developed as a part of a Laboratory Directed Research and Development (LDRD) project at NREL.

**Author contributions**

Christopher M. Caskey performed synthesis and basic characterization of the Sn-N samples, with contributions from Andriy Zakutayev. Aaron Holder, Vladan Stevanovic, Xiuwen Zhang, and Stephan Lany contributed theoretical structure prediction results. Alon Kukliansky, Amir Natan and Aaron Holder performed Raman calculations. Sarah Shulda and Svitlana Pylypenko performed XPS analysis. Ryan M. Richards, William Tumas, and David S. Ginley provided numerous useful inputs. David Biagioni helped by comparing the XRD patterns, and John D. Perkins assisted with RBS analysis. Steve Christensen, Craig P. Schwartz, Dennis Nordlund, David Prendergast, were responsible for the XAS measurements and simulations. Bernardo Orvananos, Wenhao Sun and Gerbrand Ceder provided the structure similarity calculations. David Diercks is responsible for the TEM/SAED data interpretation. The manuscript has been written by Andriy Zakutayev and Christopher M. Caskey, with contributions from Aaron Holder and all other co-authors. Andriy Zakutayev also planned and coordinated this research project.

**Synthesis of a mixed-valent tin nitride and considerations of its possible crystal structures**

Christopher M. Caskey,[1,2] Aaron Holder,[1] Sarah Shulda[2], Steve Christensen,[1] David Diercks,[2] Craig P. Schwartz,[3] David Biagioni,[1] Dennis Nordlund,[3] Alon Kukliansky,[4] Amir Natan,[4] Bernardo Orvananos,[6] Wenhao Sun,[6] Xiuwen Zhang,[7] Gerbrand Ceder,[6] David Prendergast,[5] David S. Ginley,[1] William Tumas,[1] John D. Perkins,[1] Vladan Stevanovic,[1,4] Svitlana Pylypenko,[2] Stephan Lany,[1] Ryan M. Richards,[1,2] Andriy Zakutayev[1]

1) National Renewable Energy Laboratory
2) Colorado School of Mines
3) SLAC National Accelerator Laboratory
4) Tel Aviv University
5) Lawrence Berkeley National Laboratory
6) Massachusetts Institute of Technology
7) University of Colorado Boulder

**Experimental methods**

*(a) Thin Film Synthesis*

Tin nitride films were grown by radio-frequency (RF) reactive sputtering of a 50-mm diameter tin metal target in a nitrogen and argon atmosphere. Glass substrates were placed at a 45° angle to the target producing a gradient in target-substrate distance. Temperature gradient was introduced by inserting a small piece of metal between the heater and the ¼ of the substrate, while leaving the rest of the substrate suspended in vacuum. The power supplied to the Sn target was 25 W. The flow rate of nitrogen was 10 sccm, and the flow rate of argon was also 10 sccm. The chamber pressure was 20 mTorr during depositions, and the base pressure of the chamber was <2x10$^{-6}$ Torr.

*(b) Composition and structure measurements*

The structure of the resulting Sn-N thin films were characterized by X-ray diffraction (XRD) using Cu *Kα* radiation on a Bruker D8 x-ray diffractometer with a 2D detector and automated X-Y sample stage. More detailed XRD measurements on the Sn-N powders scraped from the multiple thick Sn-N films were done on an XRD instrument with a rotating anode and point detector. The composition of the Sn-N samples was measured using Rutherford backscattering (RBS) of He atoms in the 0.4 – 2.0 MeV energy range. The samples measured by RBS were grown thin and on silicon instead of silica to allow accurate anion content determination.

*(c) Raman scattering*

Raman spectra were collected using a Renishaw inVia confocal Raman microscope configured with 532 nm laser excitation at 5% power, an 1800 mm$^{-1}$ grating, and CCD array detector. The measurements were performed using a 5× objective with a spot size of ~10 μm. To compensate for the small spot size, 6 spectra were averaged together at each measurement point.

*(d) X-ray Absorption Spectroscopy (XAS)*

Experimental X-ray absorption spectra (XAS) were taken on SnN$_{1-δ}$ samples at beamline 8-2 at the Stanford Synchrotron Radiation Lightsource (SSRL). A nominal resolution of ΔE/E of 10$^{-4}$ is provided by this beamline.[72] Data was collected both by using total electron yield (TEY) with penetration depth of ~10 nm.[73]

*(e) X-ray Photoemission Spectroscopy (XPS)*

XPS analysis was performed on a Kratos Nova X-ray photoelectron spectrometer with a monochromatic Al Kα source operated at 300 W. Survey spectra were acquired at pass energies of 160 eV with high-resolution spectra of C 1s, O 1s, Sn 3d, and N 1s at 20 eV. Data analysis was performed using CasaXPS software. A linear background was used for C, O, and N 1s spectra and a Shirley background was utilized for the Sn 3d spectra analysis. All spectra were charge referenced to C1s at 284.8. After background subtraction and charge referencing the spectra were fitted with mixed Gaussian-Lorentzian functions, constraining all component peaks FWHM to be no less than 1eV and no greater than 1.2eV. The Sn 3d high-resolution spectra exhibit two spin-orbit components corresponding to Sn 3d$_{5/2}$ (lower energy) and Sn 3d$_{3/2}$ (higher energy). For all Sn species, Sn3d$_{3/2}$ component was constrained to be shifted from the Sn $3d_{5/2}$ component by 8.4 eV and its area was set to have 0.67 times the area of the Sn $3d_{5/2}$ component. XPS data was taken from 3 areas on each sample and the reported values are averages. Peak positions and their assignments were consistent across all 3 areas for both samples and similar to those reported in literature.

*(f) Transmission electron microscopy (TEM)*

A cross-sectional TEM sample of the layered structure was prepared by typical lift-out methods in a dual-beam FEI Nova 600 FIB/SEM. Imaging and SAED measurements were performed on an FEI Titan S/TEM operated at 300kV.

*(g) Optoelectronic properties*

The optical spectroscopy was performed using a custom fiber optics spectrometer with tungeten/halogen and deuterium light sources and OceanOptics NIR and UV-VIS detectors. Temperature-dependent conductivity and room-temperature Hall effect measurements were conducted in Van der Pauw geometry using a table top system with a permanent magnet.

**Theoretical methods**

*(a) Electronic structure*

We performed first principles calculations for the structural and thermodynamic properties of the SnN structure searches using projector augmented wave (PAW) code VASP.[74] For the density functional theory (DFT) calculations, we employed the generalized gradient approximations (GGA).[75] A soft pseudopotential (PP) for N and the standard PP for Sn, that includes the $4d$ shell in the valence, was used with a 350 eV energy cutoff and a minimum k-point density of $5000/N_{at}$ for all calculations, where $N_{at}$ is the number of atoms in the unit cell. The electronic properties, band gap and absorption spectra determined from the computed dielectric function, of the top five SnN candidates were simulated using the hybrid functional HSE06, with a standard screening parameter and exact exchange fraction[76] and also using a minimum k-point density of $5000/N_{at}$. Additional details on the electronic structure methods specific to the search approaches and properties are included in their respective SI sections.

*(b) Structure prototyping search (SRS)*

Starting from the 40 $ABX_2$-type compounds as structure prototypes, tin and nitrogen atoms were substituted into the structure, and the atoms were relaxed to a local or global energy minimum. Of these 40 prototypes, three structures had distinctly lower energies. The lowest energy structure contained nitrogen-nitrogen bonds, which likely indicates phase separation. The results of the structure prototyping search method are reported in Table S1.

*(c) Genetic algorithm (GSGO)*

The global space group optimization (GSGO) method[77] starts from the unbiased guess of randomly selected lattice vectors and random atomic positions within a (super) cell as input for a sequence of ab initio calculations of total energy of locally relaxed trial structures to search for a global minimum via an evolutionary-algorithm selection. The initial GSGO structure search was performed for SnN with 4~12 atoms in the cell. The population size was set to 64 and the 16 worst individuals were replaced by offspring at each generation. The rate of crossover versus mutation was set to 0.7. Two independent GSGO runs with up to 12 generations were performed for each run.

### *(d) Random Structure Search (RSS)*

Random structures were generated for over 6,000 SnN starting configurations of up to 20 formula units. 4,000 of the random structures were generated using USPEX software[78] with constraints imposed on the interatomic distances of Sn-Sn and N-N to be > 2.0 Å to preferentially avoid metallic tin and molecular $N_2$ configurations. An electronic bias, applied by an attractive or repulsive non-local external potential (NLEP)[79] on the Sn 5s orbitals was used to influence them towards a 2+ or 4+ electronic state during the preliminary structural relaxation steps of the structure optimization calculations (in accordance with the electronic structure methods detailed in Section a).

We also generated 2,000 SnN starting configurations using random superlattices (RSL) implemented in a recently developed polymorph sampling technique.[80] The RSL structures generation/sampling is motivated by the Ab Initio Random Structure Searching method[81] and similarly starts with the random choice of unit cell parameters a, b, c, α, β, γ. To achieve predominant cation-anion coordination different types of ions are distributed randomly over two interpenetrating grids of points. Therefore, in the second step the cation and anion grids are constructed using the planes of a supperlattice constructed using a reciprocal lattice vector $G = n_1 g_1 + n_2 g_2 + n_3 g_3$ with random $n_1$, $n_2$, $n_3$, i.e., random supperlattice. The two grids are constructed by discretizing the planes corresponding to the minima (cation grid) and the maxima (anion grid) of the plane wave. In the third step the ions are distributed over the two grids. To ensure homogeneous distribution and that that no two ions of the same kind are too close, the probability distribution is constructed by placing a gaussian centered at each occupied grid point. The next ion is then placed on a grid point chosen randomly among those that have low probability. Finally, the structure is converted from fractional back into the real coordinates and the scaling factor is adjusted such that the minimal distance between any two atoms is larger than a certain threshold. The results of both RSS were filtered to remove identical structures or those that contained phase-separated Sn-metal / N-dimer regions and are displayed in Figure S9.

### *(e) Raman spectra*

We have used both spin un-polarized PBE and van der Waals (VdW) corrected PBE for the calculation of the vibrational properties of the candidate SnN structures. Before vibrational analysis the structures were fully relaxed till forces were below 0.01 eV/Å and the total energy was converged to below $10^{-6}$ eV. Vibrational calculations were performed using DFT linear response (DFPT). Raman intensities are estimated by the derivative of the dielectric susceptibility tensor with respect to the normal mode[82,83]. We have used DFPT for both the vibrations and the dielectric susceptibility and have performed the calculations using the vasp_raman.py package.[82,84] A k-point mesh of 5x5x5 was used for all structures and an energy cutoff of 500 eV was found to be sufficient – some structures were re-calculated with a cutoff of 700 eV with no significant changes. Phonon band structures were calculated using the supercell approach with the finite difference method. The Phonopy package[85] was used in order to setup and run the calculations. The path for the phonon spectrum was chosen using AFLOW[86]. For VdW interactions we used DFT-D2 method of Grimme.[87]

*(f) X-ray absorption spectra*

We calculated NEXAFS spectra using the OCEAN package,[88,89] which generates X-ray absorption spectra by solving the Bethe-Salpeter equation (BSE) within a basis of electron and hole states (and associated core-hole dielectric screening) provided by the Kohn-Sham orbitals of density functional theory (DFT).[90] The DFT electronic structure was calculated within the Perdew Berke Ernzerhof (PBE) generalized gradient approximation using the Quantum ESPRESSO code, and efficient numerical sampling of the Brillouin zone was enabled through use of the Shirley interpolation scheme.(14-17) [91,92,93,94]

*(g) Structure similarity*

The structural similarity, $\text{Sim}_{\text{struct}}$, measures the geometrical and chemical similarity between two structures. It is equal to one when two structures are identical and equal to zero when they are completely different.[95,96] The structural similarity is defined as[95]

$$\text{Sim}_{\text{struct}}(s_1, s_2) = \max_{\text{all matching}} \frac{\sum_{ss_{1,j}, ss_{2,j} \in \text{matching}} \text{Sim}_{\text{substruct}}(ss_{1,j}, ss_{2,j})}{n_{\text{lcm}}}, \quad (1)$$

where $s_i$ is the structure $i$, $ss_{i,j}$ is the substructure $j$ of structure $i$, and $\text{Sim}_{\text{substruct}}$ is the substructural similarity. Furthermore, $n_{\text{lcm}}$ is the least common multiple between the number of substructures in the two structures. $\text{Sim}_{\text{substruct}}$ is calculated following Eqs. (3), (5), and (6) from Ref. 97 using a constant c of $7.07 \times 10^{-2}$. However, we here modify the original definition of the ionic similarity[98] because we only compare structures that contain the same elements. We define it as one when two elements are equal (independently of their oxidation state) and zero otherwise.

We compute the similarity between $Sn_3N_4$ and $Sn_2N_2$-SG166, and the similarity between $Sn_3N_4$ and three other polymorphs that are more stable: $Sn_4N_4$-SG6, $Sn_8N_8$-SG62 and $Sn_4N_4$-SG12. The structural and substructural similarity depends on the similarity of the coordination environment between two substructures. [95,97] The coordination environment tends to be similar among crystals of similar chemistries.[99] By quantifying the similarity of the predicted SnN structures to the observed $Sn_3N_4$ structure, we identify the candidate SnN structure that possesses the most similar substructure coordination to $Sn_3N_4$.

**Experimental and Theoretical Results**

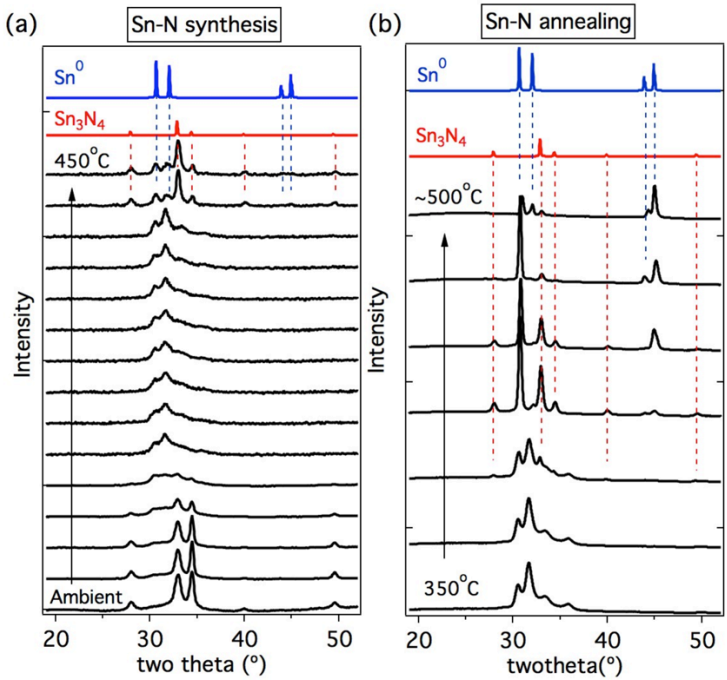

Fig S1 X-ray diffraction (XRD) results as a function of substrate temperature for (a) as synthesized (b) annealed of $SnN_{1-\delta}$ samples. The $SnN_{1-\delta}$ phase is deposited in intermediate temperature range. During annealing process, the $SnN_{1-\delta}$ phase decomposes on $Sn_3N_4$ and Sn at higher temperatures.

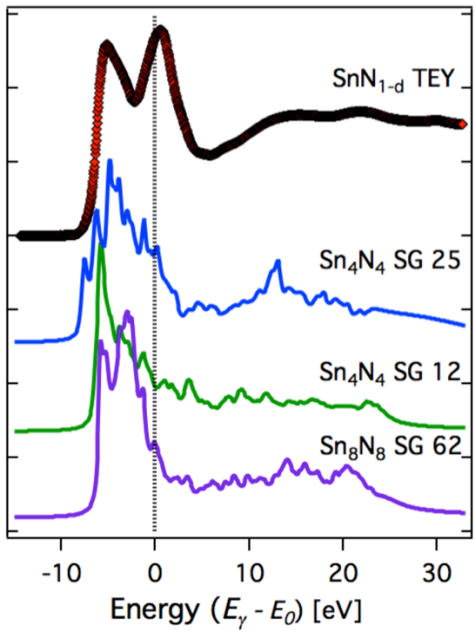

Fig. S2 Measured and simulated nitrogen 1s x-ray absorption spectra (XAS) of SnN$_{1-\delta}$ samples indicates that the SG25, SG12 and SG62 structure candidates are less likely than the SG166, SG6 and SG227 candidates shown in Fig. 9 of the main text

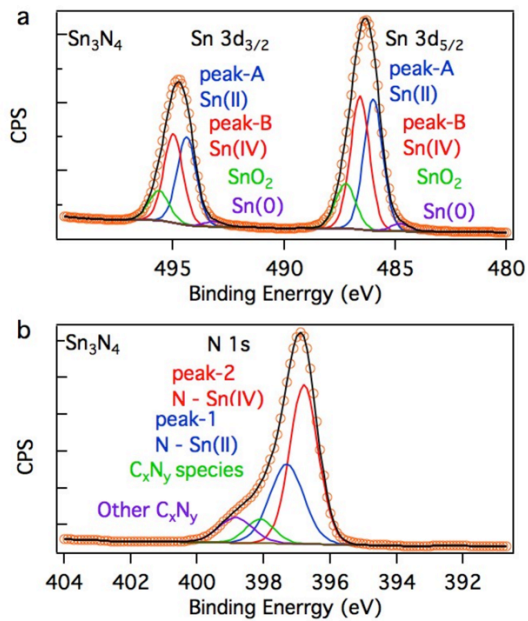

Fig. S3 X-ray photoemission spectroscopy (XPS) measurement results for Sn$_3$N$_4$ reference samples on (a) Sn and (b) N edge. The Sn(IV)/Sn(II) ratio is larger compared to that for SnN$_{1-\delta}$ samples shown in Fig. 10 of the main text

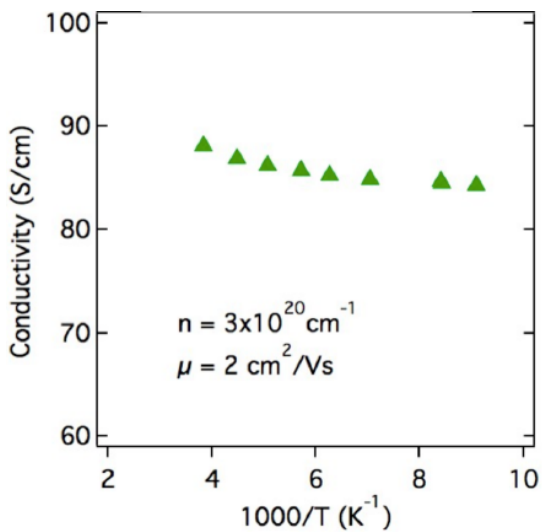

Fig. S4 Electrical conductivity as a function of temperature for SnN$_{1-\delta}$ samples shows degenerate behavior, consistent with free carrier absorption observed in experimental spectrum in Fig. 12 of the main text

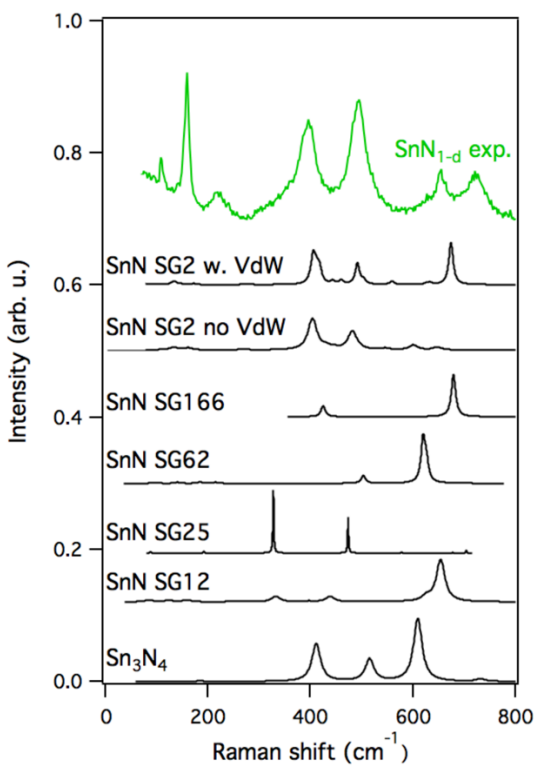

Fig. S5 Simulated Raman patterns of three likely SnN candidate structures, in comparison with the experimentally measured $SnN_{1-\delta}$ pattern and the $Sn_3N_4$ simulated reference patern

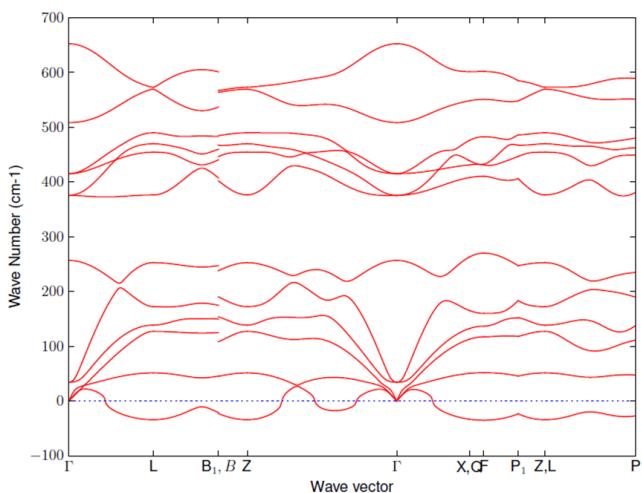

Fig. S6 Phonon analysis of the SG166 ideal cell, showing imaginary modes, which indicate dynamic instability. To relax further the structure we have built a 2x2x2 cell, performed a gamma point vibrational analysis, which revealed the same mode.

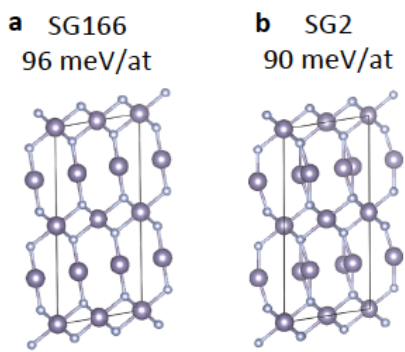

Figure S7 (a) Ideal (left, SG166) and (b) distorted (right, SG2) SnN structure, showing a broken symmetry. The relaxed structure is lower in energy by 0.15eV for the 2x2x2 super-cell, which is a very small difference.

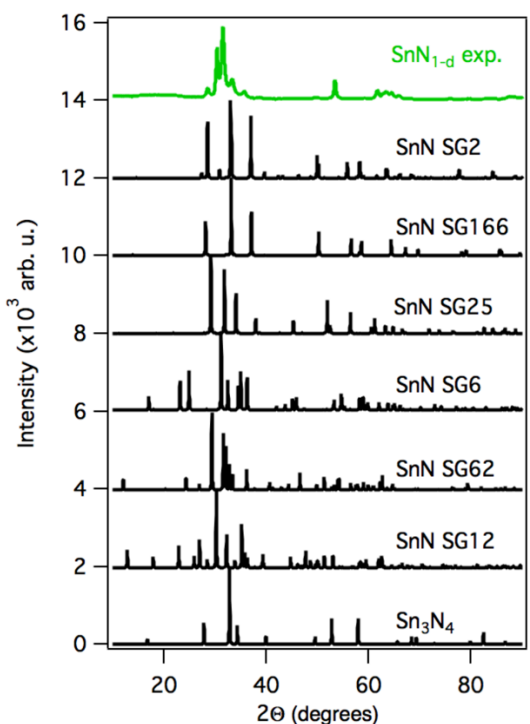

Figure S8. XRD patterns of the five most likely SnN structural candidates (SG 12, 62, 6, 25, 166 in order of increasing energy), selected based on the automated peak search-match algorithm (Fig.7 of the main text). These are shown in comparison with the experimental $SnN_{1-\delta}$ pattern, as well as the theoretical $Sn_3N_4$ and the distorted SnN SG2 reference patterns.

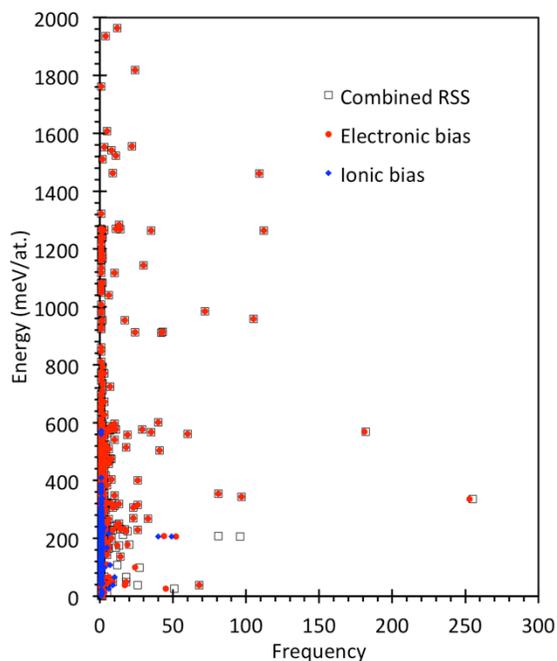

Figure S9. The results of the theoretical SnN structure search methods in the broad energy window, excluding identical and phase-separated compounds.

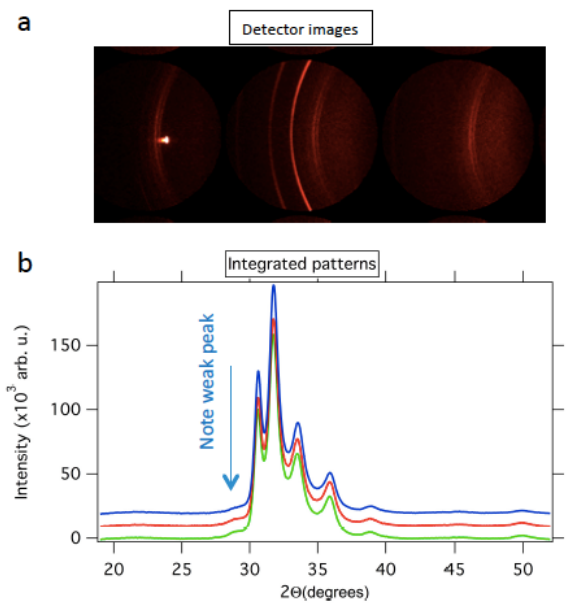

Figure S10. (a) XRD detector images and (b) integrated XRD patterns of SnN$_{1-\delta}$ samples grown at different target-substrate distances

Table S1. The results of the structure prototyping search (SRS) method for identification of the SnN structure. The energy scale is defined with respect to the energy of the above SG12 polymorph identified by other structure search methods such as RSS.

| Structure prototype | Space group | Energy (eV/atom) |
|---|---|---|
| $CuBiS_2$ | 62 | 5 |
| $CaMnSb_2$ | 62 | 13 |

Table S2 The determined d-spacing and the corresponding measured 2θ values in Cu Kα radiation for the of $SnN_{1-\delta}$ samples

| d-spacing (Å) | 2θ (°) Cu Kα | Intensity | Comments |
|---|---|---|---|
| 3.13 | 28.56 | Moderate | Very weak in films |
| 2.94 | 30.42 | Very strong | |
| 2.84 | 31.52 | Very strong | |
| 2.67 | 33.58 | Strong | |
| 2.51 | 35.74 | Moderate | |
| 2.31 | 38.98 | Weak | |
| 2.26 | 39.90 | Very weak | |
| 1.84 | 49.60 | Weak | |
| 1.75 | 52.42 | Strong | |
| 1.56 | 59.12 | Very weak | |
| 1.50 | 61.70 | Moderate | |
| 1.48 | 62.70 | Moderate | |
| 1.47 | 63.46 | Moderate | Cluster |
| 1.45 | 64.40 | Moderate | |
| 1.42 | 65.88 | Moderate | |
| 1.30 | 72.72 | Very weak | |
| 1.25 | 75.88 | Very weak | |
| 1.21 | 79.50 | Very weak | |
| 1.16 | 83.68 | Weak | |
| 1.11 | 87.62 | Moderate | Broad: likely a cluster |
| 1.04 | 95.88 | Very weak | |
| 1.01 | 99.32 | Very weak | |
| 0.99 | 102.30 | Weak | |
| 0.94 | 109.46 | Weak | |
| 0.89 | 120.14 | Very weak | |
| 0.86 | 128.14 | Very weak | |
| 0.82 | 141.02 | Weak | Very broad |